\documentclass[sigconf,screen]{acmart}
\usepackage{tabularx}

\newcommand\clearrow{\global\let\rowmac\relax}
\clearrow
 \usepackage{adjustbox}
\usepackage[flushleft]{threeparttable}
\usepackage{amsmath, amsfonts}
\usepackage{algorithmic}
\usepackage{graphicx}
\usepackage{textcomp}
\usepackage{xcolor}
\usepackage{xspace}
\usepackage{booktabs}
\usepackage{multirow}
\usepackage{subcaption}
\usepackage{enumitem}
\usepackage{graphicx}  %
\usepackage{microtype}
\usepackage{balance}
\graphicspath{ {./figures/} }
\AtBeginDocument{%
  \providecommand\BibTeX{{%
    \normalfont B\kern-0.5em{\scshape i\kern-0.25em b}\kern-0.8em\TeX}}}

\setcopyright{rightsretained}
\acmPrice{}
\acmDOI{10.1145/3597926.3598135}
\acmYear{2023}
\copyrightyear{2023}
\acmSubmissionID{issta23main-p369-p}
\acmISBN{979-8-4007-0221-1/23/07}
\acmConference[ISSTA '23]{Proceedings of the 32nd ACM SIGSOFT International Symposium on Software Testing and Analysis}{July 17--21, 2023}{Seattle, WA, USA}
\acmBooktitle{Proceedings of the 32nd ACM SIGSOFT International Symposium on Software Testing and Analysis (ISSTA '23), July 17--21, 2023, Seattle, WA, USA}
\received{2023-02-16}
\received[accepted]{2023-05-03}
\usepackage[T1]{fontenc}

\begin{document}

\title{How Effective Are Neural Networks for Fixing Security Vulnerabilities}

\author{Yi Wu}
\affiliation{%
  \institution{Purdue University}
  \city{West Lafayette}
  \country{USA}}
\email{wu1827@purdue.edu}

\author{Nan Jiang}
\affiliation{%
  \institution{Purdue University}
  \city{West Lafayette}
  \country{USA}}
\email{jiang719@purdue.edu}

\author{Hung Viet Pham}
\authornote{This work is done when Hung Viet Pham and Thibaud Lutellier were at University of Waterloo.}
\affiliation{%
  \institution{York University}
  \city{Toronto}
  \country{Canada}}
\email{hvpham@yorku.ca}

\author{Thibaud Lutellier}
\authornotemark[1]
\affiliation{%
\institution{University of Alberta}
  \city{Camrose}
    \country{Canada}}
\email{lutellie@ualberta.ca}

\author{Jordan Davis}
\affiliation{%
  \institution{Purdue University}
  \city{West Lafayette}
  \country{USA}}
\email{davi1304@purdue.edu}

\author{Lin Tan}
\affiliation{%
  \institution{Purdue University}
  \city{West Lafayette}
  \country{USA}}
\email{lintan@purdue.edu}

\author{Petr Babkin}
\affiliation{%
  \institution{J.P. Morgan AI Research}
  \city{Palo Alto}
  \country{USA}}
\email{petr.babkin@jpmorgan.com}

\author{Sameena Shah}
\affiliation{%
  \institution{J.P. Morgan AI Research}
  \city{New York}
  \country{USA}}
\email{sameena.shah@jpmchase.com}

\newcommand{\logshell}{Log4Shell\xspace}
\newcommand{\bench}{VJBench\xspace}

\newcommand{\coconut}{CoCoNuT\xspace}
\newcommand{\cure}{CURE\xspace}
\newcommand{\codit}{CODIT\xspace}
\newcommand{\recoder}{Recoder\xspace}
\newcommand{\dlfix}{DLFix\xspace}
\newcommand{\sequencer}{SequenceR\xspace}
\newcommand{\rewardrepair}{RewardRepair\xspace}
\newcommand{\transfer}{TRANSFER-PR\xspace}

\newcommand{\hopity}{Hoppity\xspace}
\newcommand{\tbar}{TBar\xspace}
\newcommand{\graphtoedit}{Graph2Edit\xspace}

\newcommand{\defects}{Defects4J\xspace}
\newcommand{\defectsold}{Defects4J v1.2\xspace}
\newcommand{\defectsnew}{Defects4J v2.0\xspace}
\newcommand{\quixbugs}{QuixBugs\xspace}

\newcommand{\javalang}{\texttt{javalang}\xspace}
\newcommand{\javaparser}{\texttt{JavaParser}\xspace}
\newcommand{\code}[1]{\texttt{\small #1}\xspace}

\newcommand{\numllms}{four} %
\newcommand{\numjavavul}{50} %
\newcommand{\newvul}{15} %
\newcommand{\newvulall}{42}
\newcommand{\newCWEsingle}{six} %
\newcommand{\newCWEall}{twelve}  %
\newcommand{\numapr}{four}

\newcommand{\todoc}[2]{{\textcolor{#1}{\textbf{#2}}}}
\newcommand{\todoblack}[1]{{\todoc{black}{\textbf{[[#1]]}}}}
\newcommand{\todored}[1]{{\todoc{red}{\textbf{[[#1]]}}}}
\newcommand{\todogreen}[1]{\todoc{green}{\textbf{[[#1]]}}}
\newcommand{\todoblue}[1]{\todoc{blue}{\textbf{[[#1]]}}}
\newcommand{\todoorange}[1]{\todoc{orange}{\textbf{[[#1]]}}}
\newcommand{\todobrown}[1]{\todoc{brown}{\textbf{[[#1]]}}}
\newcommand{\todogray}[1]{\todoc{gray}{\textbf{[[#1]]}}}
\newcommand{\todopurple}[1]{\todoc{purple}{\textbf{[[#1]]}}}
\newcommand{\todopink}[1]{\todoc{magenta}{\textbf{[[#1]]}}}
\newcommand{\todocyan}[1]{\todoc{cyan}{\textbf{[[#1]]}}}
\newcommand{\todoviolet}[1]{\todoc{violet}{\textbf{[[#1]]}}}
\newcommand{\todo}[1]{\todored{TODO: #1}}
\newcommand{\todoafter}[1]{\todoorange{TODOAFTER: #1}}

\newcommand{\lin}[1]{\todoblue{Lin: #1}}
\newcommand{\yi}[1]{\todoviolet{Yi: #1}}
\newcommand{\thibaud}[1]{\todogreen{Thibaud: #1}}
\newcommand{\hung}[1]{\todopurple{Hung: #1}}
\newcommand{\nan}[1]{\todobrown{Nan: #1}}

\renewcommand{\todoc}[2]{\relax}

\newcommand{\Comment}[1]{}

\captionsetup[figure]{font=bf,skip=6pt}%
\captionsetup[table]{font=bf,skip=6pt}%
\newcommand{\distance}{8pt}
\setlength{\textfloatsep}{6pt}%
\setlength{\floatsep}{\distance}%
\setlength{\intextsep}{\distance}%
\setlength{\dbltextfloatsep}{\distance} %
\setlength{\dblfloatsep}{\distance} %

\begin{abstract}

Security vulnerability repair is a difficult task that is in dire need of automation. Two groups of techniques have shown promise: (1) large code language models (LLMs) that have been pre-trained on source code for tasks such as code completion, and (2) automated program repair (APR) techniques that use deep learning (DL) models to automatically fix software bugs. 

This paper is the first to study and compare  Java vulnerability repair capabilities of LLMs and DL-based APR models. 
The contributions include that we (1) apply and evaluate five LLMs (Codex, CodeGen, CodeT5, PLBART and InCoder), four fine-tuned LLMs, and four DL-based APR techniques on two real-world Java vulnerability benchmarks (Vul4J and \bench), (2) design code transformations to address the training and test data overlapping threat to Codex,  (3) create a new Java vulnerability repair benchmark \bench, and its transformed version \bench-trans, to better evaluate LLMs and APR techniques, and (4) evaluate LLMs and APR techniques on the transformed vulnerabilities in \bench-trans. 
 
Our findings include that (1) existing LLMs and APR models fix very few Java vulnerabilities. Codex fixes 10.2 (20.4\%), the most  number of vulnerabilities. Many of the generated patches are uncompilable patches. (2) Fine-tuning with general APR data improves LLMs’ vulnerability-fixing capabilities. (3)  Our new \bench reveals that LLMs and APR models fail to fix many Common Weakness Enumeration (CWE) types, such as CWE-325 Missing cryptographic step and CWE-444 HTTP request smuggling. (4) Codex still fixes 8.7 transformed vulnerabilities, outperforming all the other LLMs and APR models on transformed vulnerabilities. The results call for innovations to enhance automated Java vulnerability repair such as creating larger vulnerability repair training data, tuning LLMs with such data, and applying code simplification transformation to facilitate vulnerability repair.

\end{abstract}


\maketitle

\section{Introduction}

Software  vulnerabilities, such as buffer overflows and SQL injections, have a critical impact  on global economies and can harm millions of users. Once a vulnerability is discovered, it is often crucial to fix it promptly to minimize the potential for exploitation. Yet, recent studies~\cite{li2017novel,morrison2018vulnerabilities} find that the average time to fix a vulnerability (time between the discovery and the fix) varies between 60 to 79 days,  which is still too long and provides ample opportunities for attackers to exploit these vulnerabilities. 
For example, for the severe Apache \logshell vulnerability  reported %
on November 24, 2021, the first fix was deployed by Apache 12 days after the report. During these 12 days, both Cloudflare and Cisco reported several attacks exploiting the vulnerability~\cite{hiesgen2022race}. Moreover, the initial fix proved insufficient, leaving \logshell vulnerable until a complete fix was released more than one month later. As a result, there is a need for faster vulnerability-fixing solutions.

Most vulnerability benchmarks and vulnerability repair solutions  focus either on C/C++~\cite{fu2022vulrepair, chen2022neural,huang2019using,lin2007autopag,sidiroglou2005countering,gao2021beyond,lee2018memfix,gao2015safe,muntean2019intrepair} or binaries~\cite{ma2016cdrep,avgerinos2018mayhem,musliner2015fuzzbomb,perkins2009automatically,wang2014diagnosis}. 
There is a lack of solutions and benchmarks for Java, despite it being a widely-used programming language (the third most popular language in the open-source community~\cite{github}) with \emph{many severe  vulnerabilities}. 
 
Java has been used to implement important servers, including web servers and services (e.g., Tomcat, Spring, CFX, Log4J), which are especially vulnerable to attackers. Consequently, many of the most critical vulnerabilities are in Java software. For example, Google assessed that the \logshell vulnerability in the Log4J package affected 17,000 Maven projects~\cite{log4jimpact}, and Microsoft even reported that nation-state attackers exploited the  vulnerability~\cite{log4jguide}.

Benchmarks and solutions for other programming languages often do not work or work poorly for fixing Java  vulnerabilities. 
For example, the most common vulnerabilities in C/C++ are buffer overflows~\cite{pereira2021characterizing,fan2020ac}.   
Java, as a type-safe language, is designed to avoid buffer overflows. 
 
Thus, most C/C++ techniques focusing on buffer overflow vulnerabilities are irrelevant to Java. We need new benchmarks and techniques  for fixing  Java security  vulnerabilities.

Instead of building a technique to fix Java  vulnerabilities automatically, we study and compare the space and feasibility of applying two types of techniques---learning-based automated program repair and LLMs---to fix Java security vulnerabilities automatically. 
First, learning-based program repair has gained popularity~\cite{lutellier2020coconut,jiang2021cure,zhu2021syntax,ye2022neural,chen2019sequencer,ye2022neural,hopity,zhu2021syntax,deepdebug}. These encoder-decoder approaches learn from a large  number of pairs of bugs and their fixes (in open-source projects) to fix unseen Java software bugs automatically. \emph{It would be interesting to study how effective such  learning-based program repair models  are in fixing a subset of software bugs, i.e., software vulnerabilities.}

Secondly, LLMs have recently been applied to source code~\cite{chen2021evaluating,wang2021codet5,jiang2021cure,mashhadi2021applying,prenner2022can,fan2022improving,imai2022github} and are pre-trained models that have been trained on a tremendous amount of source code (e.g., the entirety of GitHub). Different from APR models, pre-trained LLMs learn from large corpus of source code  (instead of pairs of bugs and their fixes) for various tasks such as identifier tagging and code completion. Despite learning to perform tasks different from repairing, recent study \cite{clmapr23, xia2022practical} shows that pre-trained LLMs have competitive capabilities of fixing general Java bugs \cite{just2014defects4j,quixbugs}. \emph{It would be interesting to study how effective such  LLMs  are for a different task, i.e., fixing software  vulnerabilities, when they do not see how bugs are fixed.} 

Thirdly, it would be interesting to compare deep learning (DL)-based APR techniques' and LLMs' capabilities of fixing  Java vulnerabilities. DL-based APR techniques and LLMs represent two angles of applying models for a different task. Applying DL-based APR techniques to fix vulnerabilities is using models learned from a general dataset for a specific subset of the dataset (software vulnerability is a type of software bug). Applying LLMs to fix vulnerabilities is using models learned from a different format of dataset (sequences of code) for another format (pairs of buggy and fixed code). Since LLMs do not require pairs of bugs and their fixes, LLMs are typically built from data that is orders of magnitude larger than the  training data used to train APR models. 

\emph{Would more data win or data-format matching win?}

Lastly, pre-trained LLMs are often fine-tuned to adapt to different downstream tasks~\cite{T5,codebert,graph-codebert,wang2021codet5,plbart}. A recent study \cite{clmapr23} shows that fine-tuning improves LLMs' fixing capabilities by at least 31\%. 

However, given the lack of Java vulnerability data, it is unrealistic to fine-tune LLMs for fixing Java vulnerabilities. %
Thus, \emph{it would be interesting to study how effective  LLMs fine-tuned with general APR data are in fixing software vulnerabilities.} And when compared with DL-based APR techniques, \emph{would more data plus fine-tuning win or data-format matching win?}

\subsection{Our Approach}

We conduct the first study to evaluate  and compare APR techniques' and LLMs' abilities of fixing Java vulnerabilities. 
We evaluate five LLMs (Codex~\cite{codexurl}, CodeT5~\cite{wang2021codet5}, CodeGen~\cite{nijkamp2022conversational}, PLBART~\cite{plbart} and InCoder~\cite{incoder}), four LLMs that are fined-tuned with general APR data, and {\numapr} APR techniques (CURE~\cite{jiang2021cure}, Recoder~\cite{zhu2021syntax}, RewardRepair~\cite{ye2022neural}, and KNOD~\cite{knod}) on two Java vulnerability benchmarks (Vul4J and a new \bench that we create). 
There are two main challenges.

First, there are few benchmarks available for evaluating Java vulnerability repair tools. While Vul4J~\cite{bui2022vul4j} contains 79  reproducible Java vulnerabilities, they belong to only 25 CWEs, i.e., types of vulnerabilities. 
In addition, 60\% of the CWEs in the dataset (15 types of vulnerabilities) are covered by only a single reproducible vulnerability.

To address this challenge, we develope new benchmarks. We analyze the entire National Vulnerability Database (NVD)~\cite{nvdfeed} to identify reproducible real-world Java vulnerabilities that are suitable for vulnerability repair evaluation, and use these to create our \bench benchmark. These vulnerabilities cover an additional {\newCWEall} CWE types not included by the Vul4J dataset and add more vulnerabilities to four CWE types with which Vul4J has only one vulnerability associated. The new benchmark can facilitate the evaluation of future Java vulnerability repair techniques.

The second challenge arises from the fact that Codex was trained on a substantial code corpus collected from GitHub~\cite{chen2021evaluating} and the training dataset is unreleased.
Since the projects in Vul4J and \bench are public repositories on GitHub,  one cannot be certain that the vulnerabilities in Vul4J and \bench are not in Codex's training data. 
This is a major known threat to the validity of evaluation~\cite{barz2020we, tan2015online}. 
While dataset HumanEval~\cite{chen2021evaluating}  is not in Codex's training data, 
it is for Python code completion and does not contain Java vulnerabilities. Creating new real-world benchmarks is not only expensive~\cite{just2014defects4j, bui2022vul4j}, but might also be impracticable if LLMs have been trained on all public datasets.

Our best-effort solution to mitigate this challenge is to transform the  vulnerability code in existing benchmarks. 
We use two types of code transformation: identifier renaming and code structure change. These transformations generate new equivalent programs that still retain the vulnerabilities but are not included in any open-source dataset that Codex and other LLMs may have seen. 
As a result, we create \bench-trans, a benchmark of transformed vulnerabilities, by applying two transformation strategies on vulnerabilities from Vul4J and \bench.

\subsection{Contributions}

\noindent Our paper makes the following contributions:

\begin{itemize}[leftmargin=0.4cm]
   \item We conduct the first study that evaluates the fixing capabilities of five LLMs, four fine-tuned LLMs, 
   and four APR techniques on real-world Java vulnerabilities from two benchmarks Vul4J and our new \bench. 
   Our findings include: 
    \begin{itemize}[leftmargin=0.4cm]
  \item  Existing LLMs and APR techniques fix very few Java vulnerabilities. Codex fixes 10.2 (20.4\%) vulnerabilities on average, exhibiting the best fixing
capability.  (Section~\ref{sec:rq1})
    \item  Fine-tuning with general APR data improves LLMs’ vulnerability-fixing capabilities. Fine-tuned InCoder fixes 9 vulnerabilities, exhibiting competitive fixing capability to Codex’s. (Section~\ref{sec:rq1})
   \item Codex has the highest compilation rate of 79.7\%. 
Other LLMs (fine-tuned or not) and APR techniques have low compilation rates (the lowest being 6.4\% with CodeT5 and the rest between 24.5\% to 65.2\%), showing a lack of syntax domain knowledge. (Section~\ref{sec:rq1})
    \item  LLMs and APR models, except Codex, only fix vulnerabilities that require simple changes, such as  a single deletion or variable/method replacement. (Section~\ref{sec:rq2})
    \item Our new \bench reveals that LLMs and APR models fail to fix many CWE types including CWE-172 Encoding error, CWE-325  Missing cryptographic step, 
CWE-444 HTTP request smuggling, CWE-668 Exposure of resource to wrong sphere, and CWE-1295 Debug messages revealing unnecessary information. (Section~\ref{sec:rq2})

   \end{itemize}

    \item We create two Java vulnerability benchmarks for automated program repair: (1) \emph{\bench}, which contains  {\newvulall}  reproducible real-world Java vulnerabilities that cover {\newCWEall} new CWE types, and (2) \emph{\bench-trans}, which contains 150  transformed Java vulnerabilities.

   \item We  use code transformations to mitigate the threat that 
   LLMs and black-box Codex may have seen the evaluated benchmarks.
  \item We evaluate LLMs and APR techniques' fixing capabilities on transformed vulnerabilities (\bench-trans).
 \begin{itemize}[leftmargin=0.4cm]
    \item Code transformations make LLMs and APR techniques fix fewer number of vulnerabilities.  Some models such as Codex and fine-tuned CodeT5 are more robust to code transformations. On the other hand, some transformations make the vulnerabilities easier to fix.
    (Section~\ref{sec:rq3}) 
 \end{itemize}

  \item We provide implications and suggestions for future directions (Section~\ref{sec:results}).
\end{itemize}

\section{%
New Benchmark of Java Vulnerabilities}
\label{dataset}

A Java APR benchmark must contain reproducible Java vulnerabilities with test cases exposing the vulnerabilities.  
While there is an abundance of such benchmarks for Java bugs, including Defects4J~\cite{just2014defects4j}, 
QuixBugs~\cite{quixbugs}, Bugs.jar~\cite{bugsjar}, and Bears~\cite{bears}, the only Java vulnerability benchmark for APR is Vul4J~\cite{bui2022vul4j}. 
Vul4J contains 79 vulnerabilities from 51 projects covering 25 CWE types. 
Despite a valuable first step, Vul4J offers limited coverage of CWE categories as 
explained in Introduction. %
In addition, only 35 of these vulnerabilities are applicable 
for evaluating state-of-the-art learning-based APR systems ~\cite{jiang2021cure,zhu2021syntax,ye2022neural} since these APR models only fix single-hunk bugs. Specifically, 39 of the 79 vulnerabilities are single-hunk. We can only reproduce 35 of the 39 vulnerabilities, as two bugs fail to compile, and two bugs are not reproducible with the Docker container provided by the Vul4J authors.

To extend this benchmark, we collect Java vulnerabilities following prior work~\cite{just2014defects4j}: i) The vulnerability should only be related to Java source code, ii) The fixing commit should contain at least one test case that passes on $V_{fix}$ but fails on $V_{bug}$, iii) The fixing patch should only include changes that fix the vulnerability and should not introduce unrelated changes such as features or refactoring, and iv) the vulnerability is not already in Vul4J.  %

We download all available vulnerability data in JSON format on May 13, 2022 from NVD. 
We parse this data and obtain a list of 7,116 GitHub projects by collecting the reference URLs of these %
vulnerabilities. 
We exclude projects which have less than 50\% of their code in Java, resulting in 400 Java projects containing 933 unique vulnerabilities. We then try to identify the fixing commits for each of the 933 vulnerabilities by manually checking the reference links provided in the vulnerability report or by searching the vulnerability ID in the GitHub repository if no link is provided. We find vulnerability-fixing commits for 698 vulnerabilities. Then we manually filter out 185 vulnerabilities whose fixing commits contain non-Java changes and 314 vulnerabilities that do not have test cases in their fixing commits. We now have 199 vulnerabilities, each with test cases and a corresponding Java-only fixing commit. We then successfully reproduce {\newvulall} Java vulnerabilities that are not included in Vul4J, using building tools such as Maven or Gradle.

\begin{figure}[!t]
   \centering
   \includegraphics[width=0.95\columnwidth]{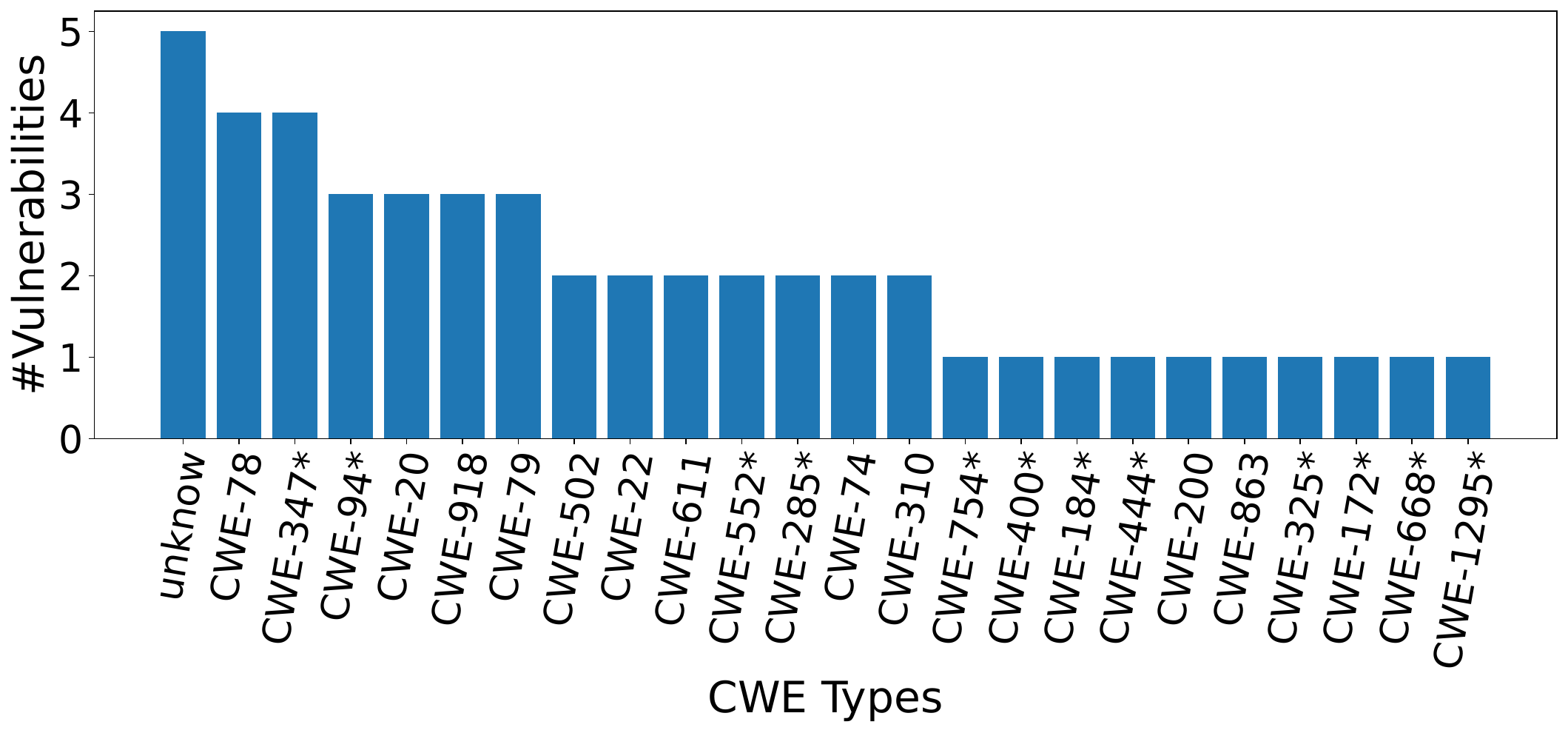}
   \caption{ CWE Type Distribution of VJBench (* denotes the new CWE types not included in Vul4J).}
   \label{fig:vjbench_distribution}
  
\end{figure}

We end up with a dataset of \textbf{{\newvulall} new reproducible real-world Java vulnerabilities} from thirty open-source projects.  In detail, our dataset consists of \emph{27 multi-hunk vulnerabilities} from twenty-two projects and \emph{{\newvul} single-hunk vulnerabilities} from eleven projects. As Figure~\ref{fig:vjbench_distribution} shows, these 42 vulnerabilities covers a total of 23 CWE types. Furthermore, our dataset introduces \textbf{12 new CWE types} (denoted by * in Figure~\ref{fig:vjbench_distribution}) not included in Vul4J and supplements four CWE types (CWE-78, CWE-200, CWE-310, CWE-863) for which Vul4J only has one example. 

 Table~\ref{vul_table} describes the {\newvul} new single-hunk vulnerabilities of twelve CWE types in our \emph{\bench} benchmark. 
There are six new unique CWE types of vulnerabilities not present in Vul4J. 
As a result, there are {\newvul} vulnerabilities from \bench and 35 vulnerabilities from Vul4J, a total of \textbf{50 vulnerabilities} that we use in our study.

\begin{table}[t]
\footnotesize
\centering
\caption{List of the {\newvul} new single-hunk vulnerabilities categorized by their corresponding CWE. The vulnerability IDs compose of the project name and the bug index.
* denotes the {\newCWEsingle} new CWE types that our benchmark adds compared to Vul4J. Jenkins-1 and Flow-2 both belong to two CWE categories. }
\label{vul_table} 

\begin{tabular}{r@{}lll}
\toprule
\multicolumn{2}{l}{\textbf{CWE}} & \textbf{Description} & \textbf{Vulnerability IDs}\\ 
\midrule
20   & & Improper Input Validation  & Pulsar-1 \\
22   & & Improper limitation of path name & Halo-1 \\
    & & to a restricted directory & \\
74   & & Improper Neutralization of Elements & Ratpack-1 \\
    & & in Output ('Injection')  & \\
79   & & Cross-site Scripting & Json-sanitizer-1 \\
172  &* & Encoding error & Flow-1  \\
200  & & Exposure of sensitive information & Jenkins-1, Jenkins-2, Jenkins-3 \\
325  &* & Missing cryptographic step & Jenkins-1\\
347  &* & Improper Verification of  & BC-Java-1\\
    & & Cryptographic Signature & \\
444  &* & HTTP request smuggling & Netty-1, Netty-2 \\
611  & & Improper restriction of XML external & Quartz-1, Retrofit-1 \\
    & & entity reference & \\
668  &* & Exposure of resource to wrong sphere & Flow-2 \\
1295 &* & Debug messages revealing & Flow-2 \\
    & & unnecessary information & \\
\textit{unk} & & no specific CWE category & Jinjava-1\\
\bottomrule
\end{tabular}

\end{table}

\section{Large Language Models and APR Techniques}
\label{models}

\subsection{Large Language Models}
\label{llm}

We select five LLMs, i.e., Codex, PLBART, CodeT5, CodeGen and InCoder, because they are (1) state-of-the-art, (2) capable of performing code generation tasks without any modifications to the models or additional components (e.g., CodeBERT~\cite{codebert} GraphCodeBERT~\cite{graph-codebert} are excluded),  and (3)  trained with enough source code so that they can understand code to some extent (e.g., we exclude T5~\cite{T5},  GPT-2~\cite{gpt-2}, GPT-Neo~\cite{gpt-neo} and GPT-J~\cite{gpt-j}, whose training data is over 90\%  text). In this work, we study the LLMs in two settings: as is and fine-tuned with general APR data. 

\subsubsection{Large Language Models As Is}
In this section, we introduce the details of the studied LLMs and how to use them for fixing vulnerabilities. Table~\ref{tab:model_size} provides the model sizes and their training data information.

\begin{table}[t]
\begin{center}
\footnotesize

\caption{Input Formats of Large Language  Models}
\label{tab:prmopt}
\begin{tabular}{ll}

\toprule
\textbf{Model} & \textbf{Input Format }\\
\midrule
\multirow{3}{*}{Codex} & Comment  buggy lines (BL) with hint ``BUG:'' and ``FIXED:'' \\ %
& Prefix prompt: Beginning of the buggy function to  BL comment \\
& Suffix prompt: Line after BL comment to end of the buggy function\\
\midrule
CodeT5      & Mask  buggy lines with $<$extra\_id\_0$>$ and  input the  buggy function \\
\midrule
CodeGen     & Input beginning of the buggy method to line before buggy lines \\
\midrule
PLBART      & Mask  buggy lines with $<$mask$>$ and  input the  buggy function \\
\midrule
InCoder     & Mask  buggy lines with $<$mask$>$ and  input the  buggy function \\
\midrule
Tuned LLMs & Comment  buggy lines and  input the  buggy function \\

\bottomrule
\end{tabular}
    
\end{center}
\end{table}

\begin{figure}[!t]
   \centering
   \includegraphics[width=1\linewidth]{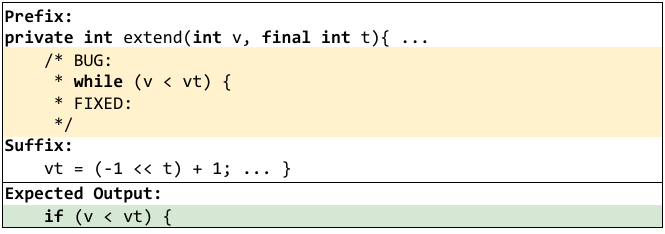}
   \caption{An example input to Codex and its expected output}
   \label{fig:input_codex}
\end{figure}

\smallskip\noindent\textbf{Codex~\cite{chen2021evaluating}:}
Codex is a GPT-3-based~\cite{chen2021evaluating,gpt-3} language model with 12B parameters trained on both natural language and source code. %
We use the davinci-002 model (as of July 2022), which is supposed to be %
the most accurate  Codex model~\cite{codexurl}.
We focus on Codex's insertion mode as it provided the best results in our preliminary study among the three main modes: completion, insertion, and edit. 

\smallskip\noindent\textbf{CodeT5~\cite{wang2021codet5}:}
CodeT5 is an encoder-decoder transformer model~\cite{attention} pre-trained with an identifier-aware denoising objective and with 
 bimodal dual generation tasks.
It is trained on a corpus of 5.2 million code functions and 8.3 million natural language sentences from open-source repositories in six programming languages including Java. In this work, we use the largest CodeT5 model released, which has 770M parameters.

\smallskip\noindent\textbf{CodeGen~\cite{nijkamp2022conversational}:} CodeGen models are a series of autoregressive decoder-only transformers trained for conversational program synthesis. Their training data consists of 354.7B natural language tokens from THEPILE dataset and 150.8B programming language tokens extracted from a subset of the Google BigQuery database. In this work, we apply the CodeGen model which contains 6B parameters (the larger model with 16B parameters is not used due to the limitation of our machine).

\smallskip\noindent\textbf{PLBART~\cite{plbart}:}
PLBART uses an encoder-decoder transformer architecture with an additional normalization layer on the encoder and decoder. It's pre-trained on functions extracted from Java and Python GitHub repositories via denoising autoencoding. %
Two PLBART models of different sizes are available, and we use the larger model containing 400M parameters.

\smallskip\noindent\textbf{InCoder~\cite{incoder}:}  InCoder models follow XGLM~\cite{xglm}'s decoder-only architecture and are pre-trained on the masked span prediction task. Its pre-training data comes from open-sourced projects on GitHub and GitLab, and StackOverflow posts. %
There are two InCoder models of different sizes released, and we use the larger one which contains 6B parameters.

\smallskip\noindent\textbf{Input Formats:}
Table~\ref{tab:prmopt} illustrates the input format we used for each model. For Codex, we adopt an input format similar to the one used in prior work~\cite{pearce2022examining}. The prompt includes the commented buggy code with hint words ``BUG:" and ``FIXED:" to signify the location of the bug and to guide Codex towards generating a fixed version of the code. If the number of input tokens exceeds the maximum number for a model, we truncate the code and input the code around the buggy lines. Since it is unclear how the commented buggy line prompts will affect the models' fixing capabilities, we experiment with the input with and without commented buggy lines for each model. \autoref{fig:input_codex} shows an example of the input and expected output of Codex with buggy lines commented by \code{/* BUG .. FIXED */}.

\begin{table}[]
    \centering
        \caption{Model size (number of parameters) and training data size of the five LLMs we apply and report in this work}
    \label{tab:model_size}
 \begin{adjustbox}{width=\columnwidth,center}
    \begin{tabular}{l@{\quad}lrrrrr}
    \toprule
        \multicolumn{2}{l}{ %
        } 
        & \textbf{Codex} & \textbf{CodeT5} & \textbf{CodeGen} & \textbf{PLBART}& \textbf{InCoder} \\
    \midrule
        \multicolumn{2}{l}{\#Parameters} & 12B & 770M & 6B & 400M& 6B \\ 
    \midrule
        Training Data  & NL & 45.0TB & - & 1.1TB & 79.0GB& 57.0GB \\
        Raw Size & PL & 159.0GB & - & 436.3GB & 576.0GB& 159.0GB \\
    \midrule
           Training Data %
           & NL & 499.0B & - & 354.7B & 6.7B & -\\
      \#Tokens  & PL & 100.0B & - & 150.8B & 64.4B& - \\
    \midrule
        Training Data & NL & - & 5.2M & - & 47.0M& - \\
       \#Instances & PL & - & 8.3M & - & 680.0M & -\\
    \bottomrule
    \end{tabular}
       \end{adjustbox}

\end{table}

\subsubsection{Fine-Tuned Large Language Models}
We also study the fixing capabilities of fine-tuned LLMs, since fine-tuning is a common technique to adapt a pre-trained LLM to a specific downstream task, such as code summarization or code translation~\cite{wang2021codet5,codebert,incoder,T5}. However, due to the lack of vulnerabilities as fine-tuning data, we use the LLMs fine-tuned with general APR data, shared by existing work~\cite{clmapr23}. Prior work~\cite{clmapr23} fine-tuned LLMs with a training dataset containing 143,666 instances collected from open-source GitHub Java projects~\cite{zhu2021syntax}. Each data instance is a pair of buggy code and fixed code. In detail, \cite{clmapr23} used the Adam optimizer with a learning rate of $1e^{-5}$, set batch size to one and fine-tuned for one epoch. The fine-tuned LLMs are supposed to be adjusted to vulnerability fixing task to some extent due to the similarity between vulnerability fixing and general bug fixing. We perform a search and confirm that none of the vulnerabilities we study in this work is present in the APR training data used to fine-tune the LLMs.

We cannot fine-tune Codex, since it does not offer any fine-tuning API and there is also no fine-tuned Codex available. The last row of Table~\ref{tab:prmopt} describes the input format for using fine-tuned LLMs, where the buggy lines are given as commented lines, and the entire function is input into the fine-tuned LLMs to generate the patched lines~\cite{clmapr23}.

\subsection{APR Techniques}
\label{apr}

We select four state-of-the-art learning-based APR techniques trained for Java bugs. These APR techniques  need to be open-sourced so that we can run them on our new vulnerability benchmarks.

\smallskip\noindent\textbf{CURE~\cite{jiang2021cure}} applies a small language model (pre-trained with 4.04M code instances) to the CoCoNuT's~\cite{lutellier2020coconut} encoder-decoder architecture to learn code syntax and propose a new code-aware strategy to remove invalid identifiers and increase the compilation rate during inference. CURE is trained with 2.72M APR  instances.

\smallskip\noindent\textbf{Recoder~\cite{zhu2021syntax}} uses an tree-based deep learning network that is trained on 82.87K APR training instances. It focuses on generating edits to modify buggy ASTs to form the patched ASTs.

\smallskip\noindent\textbf{RewardRepair~\cite{ye2022neural}} includes compilation in the calculation of the model's loss function  to increase the number of compilable (and correct) patches. This is different from CURE as the loss function increases the number of compilable patches during training. Overall, RewardRepair is trained with 3.51M APR training instances.

\smallskip\noindent\textbf{KNOD~\cite{knod}} proposes a novel three-stage tree decoder to generate the patched ASTs, and also uses domain-knowledge distillation to modify the loss function to let the models learn code syntax and semantics. KNOD is trained with 576K APR training instances, and is the state-of-the-art DL-based APR techniques.

\section{Code Transformation}
\label{obf}

\begin{figure}[!t]
  \centering

  \begin{subfigure}[a]{0.49\textwidth}
   \includegraphics[trim=0mm 2mm 0mm 0mm,clip,width=1\linewidth]{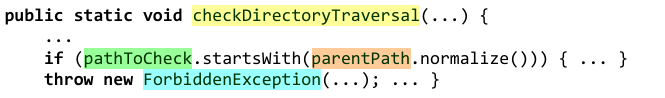} 
   \caption{Before identifier renaming}
   \label{fig:ir_before} 
  \end{subfigure}

  \begin{subfigure}[b]{0.49\textwidth}
   \vspace{2mm}
   \includegraphics[trim=0mm 2mm 0mm 0mm,clip,width=1\linewidth]{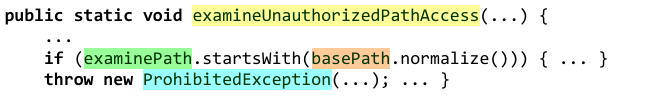} 
   \caption{After identifier renaming}
   \label{fig:ir_after}
  \end{subfigure}
  
  \caption{%
  Identifier renaming for Halo-1. Functions "startsWith" and "normalize" remain intact as they are Java library functions.
  }
  \label{fig:ir_trans}
\end{figure}

\begin{figure}
  \centering
  
  \begin{subfigure}[b]{0.49\textwidth}
   \includegraphics[trim=0mm 2mm 0mm 0mm,clip,width=1\linewidth]{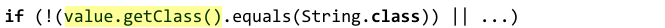} 
   \caption{Before function chaining} %
   \label{fig:mc_before} 
  \end{subfigure}

  \begin{subfigure}[c]{0.49\textwidth}
   \vspace{2mm}
   \includegraphics[trim=0mm 1mm 0mm 0mm,clip,width=1\linewidth]{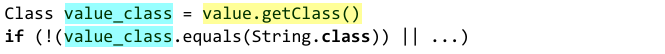} 
   \caption{After function chaining} %
   \label{fig:mc_after}
  \end{subfigure}
  
  \caption{%
  Function chaining %
  for VUL4J-30
  }
  \label{fig:mc_trans}
\end{figure}

\begin{figure}[!t]
\centering
\begin{subfigure}[a]{0.49\textwidth}
   \includegraphics[trim=0mm 2mm 0mm 0mm,clip,width=1\linewidth]{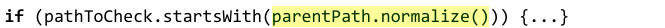}
   \caption{Before function-argument passing} %
   \label{fig:fa_before} 
\end{subfigure}

\begin{subfigure}[b]{0.49\textwidth}
   \vspace{2mm}
   \includegraphics[trim=0mm 1mm 0mm 0mm,clip,width=1\linewidth]{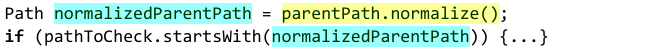}
   \caption{After function-argument passing} 
   \label{fig:fa_after} 
\end{subfigure}

\caption{%
Function-argument passing %
for Halo-1.}
 \label{fig:fa_trans}
\end{figure}

To address the challenge of training-testing data overlap, we need to create  vulnerabilities and their fixes that have not been seen by existing LLMs or APR techniques. We generate unseen vulnerabilities by transforming existing vulnerabilities to their semantically equivalent forms. %
None of the APR models and LLMs, including Codex, have seen these transformed buggy code and the corresponding fixes in their training set. 
We apply two categories of transformations to Vul4J and \bench, which are described below:

\smallskip\noindent\textbf{(1) Identifier Renaming:}
To prevent LLMs and APR models from simply memorizing the exact correct patches associated with identifier names,
we rename identifiers in the buggy code and the corresponding fixed code. 
All variables, functions, and classes defined in the project are renamed using synonyms for the original identifier names according to Java specifications. 
We use synonyms to keep the word meaning of the original identifiers. 
We do not rename identifiers from external libraries or default Java class libraries, since one often cannot modify external libraries. 
Figure~\ref{fig:ir_trans} shows an example of identifier renaming for Halo-1.

We first use the tool src2abs~\cite{srcabsgit} to extract all variable, function, and class names in the buggy function, and filter out those identifiers from Java or third-party libraries. 
 We tokenize each identifier based on camel case or snake case conventions, then use NLTK WordNet~\cite{wordnet} to generate synonyms for each word. After that, we reassemble these synonyms to form a complete identifier. We manually review and adjust the synonyms to ensure they fit the code context. Since some APR techniques need to extract identifiers from the whole project, we rename the identifiers used in the buggy function across the entire project.

\smallskip\noindent\textbf{(2) Code Structure Change:} 
We define six transformation rules to change code structures. 

\begin{itemize}[leftmargin=0.4cm]

\item\smallskip\noindent\textbf{If-condition flipping:}  negates an if-condition and swaps the code blocks in the \code{if} and \code{else} branches. %

\item\smallskip\noindent\textbf{Loop transformation:} converts a \code{for} loop to a \code{while} loop and vice versa.

\item\smallskip\noindent\textbf{Conditional-statement transformation:} turns a ternary expression (\code{var = cond ? exprTrue: exprFalse;}) 
into an \code{if-else} statement (\code{if (cond) \{var = exprTrue;\} else \{var = exprFalse;\}}), and transform a switch statement into multiple if and elseif statements, and vice versa. 

\item\smallskip\noindent\textbf{Function chaining:}  %
merges multiple function invocations into one call chain, or conversely splits a function call chain into separate function invocations. Figure~\ref{fig:mc_trans} shows an example where \code{value.getClass().equals(...);} is split into \code{Class value\_class = value.getClass();} and \code{value\_class.equals(...);}.

\item\smallskip\noindent\textbf{Function-argument passing:} %
If a locally defined variable or object is only used as a function argument, we replace the function argument with its definition statement, or we extract the function call that is passed as a function argument into a separate variable/object definition. Figure~\ref{fig:fa_trans} shows an example where the argument \code{parentPath.normalize()} is extracted and declared as a local object \code{normalizedParentPath}.

\item\smallskip\noindent\textbf{Code-order change:} alters the order of  statements if changing the order does not affect the execution results. For example, \code{ funcA(); int n =0;} can be transformed into  \code{ int n = 0; funcA();} as invoking \code{funcA()} and declaring \code{int n} do not affect each other.  

\end{itemize}

For code structure change, we manually transform the buggy function. For each buggy function, we apply all applicable transformations at once. We further confirm the equivalence of the transformed bug by reproducing them using the same test set and applying semantically equivalent patches to pass the tests.

\smallskip\noindent\textbf{A new benchmark (\bench-trans):}
In summary, to create bugs and patches that LLMs have not seen in their training set, we apply three sets of transformations (identifier renaming only, code structure change only, and both at the same time) to \bench and Vul4J, 
and create \emph{\bench-trans} that contains 3$\times$50 = 150 transformed Java vulnerabilities. 
We search in GitHub and Google the transformed code, and find no public code that is the same as the transformed buggy function. 

\smallskip\noindent\textbf{Recover patches for evaluation:} The transformed code is still realistic and human-readable. However, for the ease of evaluating the correctness of plausible patches, we maintain a dictionary that stores the mapping between the renamed identifiers and their original names. For each vulnerability, we also write a patched program for its code structure transformed version, providing a reference for future dataset users.

\section{Experiment Setup}
\label{experiment}
\begin{figure}
  \centering
    \includegraphics[width=0.95\columnwidth]{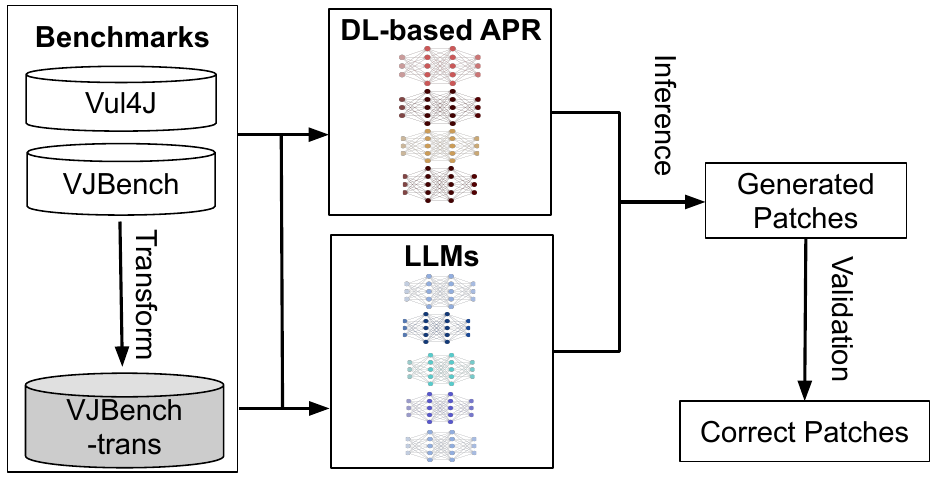} 
  \caption{Overview of our study 
  }
  \label{fig:overview}
\end{figure}

Figure~\ref{fig:overview} provides an overview of our study. First, we build a new dataset of vulnerabilities, \bench, that contains {\newvulall} new vulnerabilities. We use this new dataset and the original dataset (Vul4J) to benchmark the vulnerability-fixing capabilities of DL-based APR techniques, LLMs and fine-tuned LLMs. Each language model generates 10 patches for each bug through inference. For each APR model, we use its default beam search size and validate its top 10 patches. The generated patches are then validated using test cases and manual verification of all the patches that pass the test cases. Then, we apply code transformations on Vul4J and \bench to generate \bench-trans. Finally, we evaluate the impact of code transformations on the vulnerability-repair capabilities of all the LLMs, fine-tuned LLMs and APR techniques.

\subsection{Dataset}

In this work, we focus on fixing single-hunk Java vulnerabilities as state-of-the-art DL-based APR models are designed to fix single-hunk bugs. We filter and obtain 35 single-hunk bugs from Vul4J dataset. Along with the 15 single-hunk vulnerabilities from \bench, we have a total of 50 Java vulnerabilities. We use the perfect fault localization for these Java vulnerabilities, that is, we use the code lines that are modified in the developers' patches as the buggy lines.

\subsection{Large Language Model Setups}
We evaluate each LLM with two input setups: (1) the buggy lines are commented as part of the input and (2) without the buggy lines. We observe that InCoder fixes more vulnerabilites when the input contains buggy line comments, while the other LLMs perform better without buggy lines. We then report the best-performing setup for each model in the rest of this paper. For fine-tuned LLMs, we follow the input format with buggy line comments used in \cite{clmapr23} which is described in Table~\ref{tab:prmopt}.  %

We configure each model to generate 10 patches for each vulnerability.
For CodeT5, CodeGen, PLBART and InCoder, we set their beam search size to 10. For Codex, we set its parameter \textit{n}, the number of candidates to generate, to 10. Considering the inherent randomness of the sampling method adopted by Codex, we run it %
twenty-five times for each vulnerability to obtain the average results. We run twenty-five times to control the margin of error small ($\leq$0.3) at 95\% confidence level. We set the sampling temperature of Codex to 0.6, which is shown to have the best performance when sampling ten candidates in prior  work~\cite{chen2021evaluating}. We set the max number of newly generated tokens to 400 for Codex due to its request rate limit, and to 512 for all other LLMs.

\subsection{Patch Validation}
\label{sec:validation}

Codex insertion mode generates code to be inserted between the prefix prompt and the suffix prompt. Since we use the code before and including the buggy line comment as its prefix prompt and the code after the buggy line comment as its suffix prompt, we replace the original buggy code with the code that Codex generates. Similarly, CodeT5 generates code to replace the masked label in its input. PLBART generates the entire patched function that replaces the whole buggy function. CodeGen and InCoder are completion models that generate code to complete the given prefix prompt. We take the first complete function CodeGen and InCoder generate to replace the original buggy function. For all the fine-tuned LLMs, the fine-tuned CodeT5, CodeGen, PLBART and InCoder directly generate the patched code to replace the buggy code.

For each LLM and APR techniques, we first validate the top-10 patches they generate using the test cases from the project. Following prior work~\cite{lutellier2020coconut,jiang2021cure,zhu2021syntax,ye2022neural}, \emph{plausible patches} are patches that  pass all  test cases, while \emph{correct patches} are semantically equivalent to developer patches, and \emph{over-fitted patches} 
are patches that pass all test cases but are incorrect. We manually inspect each plausible patch to identify if it is a correct patch.

\section{Results and Findings}
\label{sec:results}
We evaluate the vulnerability fixing capabilities of five LLMs, four fine-tuned LLMs and four DL-based APR techniques on two  real-world Java vulnerability benchmarks. 

\subsection{%
RQ1: Vulnerability Fixing Capabilities} %
\label{sec:rq1}

We run Codex twenty-five times and report the average number of fixed vulnerabilities with the margin of error, because Codex's patch generation is non-deterministic. For other LLMs, we only run them once since their patch generation is deterministic (Section~\ref{experiment}).

Table~\ref{tab:repair_result} shows the fixing capabilities, i.e., the number of vulnerabilities that each approach fixes correctly, of five LLMs, four fine-tuned LLMs and four APR models. We consider the top ten patches since a recent study shows that almost all developers are only willing to examine ten patches at most~\cite{noller2022trust}. 
Results in Table~\ref{tab:repair_result} are reported as X/Y, where X is the number of vulnerabilities correctly fixed by each technique and Y is the number of vulnerabilities that are plausibly fixed.  
A vulnerability is plausibly fixed by a model if the model generates a plausible patch (definition in Section~\ref{sec:validation}).

\begin{table*}[t]
\caption{Comparison of LLMs and APR models on fixing Java vulnerabilities.  
For x/y in a cell, x denotes the number of correctly-fixed bugs, and y is  plausibly-fixed bugs (with at least one patch that passes the test cases). RewardR is RewardRepair.} 
\label{tab:repair_result}
\begin{center}
\footnotesize
\begin{tabular}{lr@{\quad}r@{\quad}r@{\quad}r@{\quad}r@{\quad}r@{\quad}r@{\quad}r@{\quad}r@{\quad}r@{\quad}r@{\quad}r@{\quad}r}
\toprule
& \multicolumn{5}{c}{\textbf{LLMs}} & \multicolumn{4}{c}{\textbf{Fine-Tuned LLMs}}  & \multicolumn{4}{c}{\textbf{APR models}}  \\
\cmidrule(lr){2-6}\cmidrule(lr){7-10}\cmidrule(lr){11-14}    

    & Codex & CodeT5 & CodeGen & PLBART & InCoder & CodeT5 & CodeGen & PLBART & InCoder & CURE & Recoder & RewardR & KNOD
    \\

\midrule
\bench (15) & \textbf{4.0}/ 4.6 & {0/0} & {1/2} & {2/3} & {2/2} & {3/4}& {3/4}& {2/3}&{3/4} & {0/1} & {1/2} & {2/3} &  0/0 \\
Vul4J  (35) & \textbf{6.2}/ 10.9 & {2/2} & {1/6} & {0/4} &{3/4}& {2/7}& {5/8}& {2/6}& {6/9}& {1/4} & {0/4} & {0/2}  &   1/1\\ 
\midrule
\textbf{Total (50)} &\textbf{10.2}/ 15.5 & {2/2} & {2/8} & {2/7} &{5/6}& {5/11}& {8/12}& {4/9}& {9/13}& {1/5} & {1/6} & {2/5} & 1/1\\ 
\midrule
Compilation Rate (\%) & 79.7 & {6.4} & {35.8} & {47.8} &{65.2}& {46.8}& {47.2}& {45.2}& {55.2}& {24.5} & {57.6} & {37.7} & {37.3}  \\ 
\bottomrule

\end{tabular}
\end{center}

\end{table*}

\subsubsection{LLMs vs. APR Techniques}

We first compare using LLMs as is with APR techniques. Here, \textit{LLMs as is} refers to that we apply Codex and LLMs under zero-shot learning and without fine-tuning.
Our results show that Codex exhibits the best fixing capability. Out of a total of 50 vulnerabilities in Vul4J and \bench,  Codex fixes an average of 10.2 vulnerabilities with a margin of error of 0.3 (at 95\% confidence). InCoder demonstrates the second best capability, fixing 5 vulnerabilities. The other LLMs and DL-based APR techniques only fix very few vulnerabilities.
Overall, LLMs and APR techniques show very limited vulnerability fixing capabilities. 

Our finding of Codex performing the best on fixing Java vulnerabilities is consistent with Codex's superior performance in repairing general bugs~\cite{xia2022practical} and in other domains~\cite{codexurl,chen2021evaluating,pearce2022examining,finnie2022robots}, possibly due to its significantly larger model size and training data size as indicated in \autoref{tab:model_size}. Our result is also consistent with recent work ~\cite{clmapr23} in showing that LLMs without fine-tuning have competitive fixing capabilities -- InCoder fix three more vulnerabilities than the best APR technique (RewardRepair). However, while ~\cite{clmapr23} shows that CodeGen, PLBART and InCoder as is can fix 18\%-23\% 
 general bug of Java APR benchmarks,  our result shows that they can fix only 4\%(2/50)-10\%(5/50) vulnerabilities of Vul4J and \bench.  In real-world, only about 1\textasciitilde 7\% of bugs are vulnerabilities, resulting in few data for models to learn from. This means that, for neural networks, fixing vulnerabilities is more difficult than general bugs and requires more domain-specific knowledge.

\smallskip\noindent\fbox{
\centering
\parbox{0.95\linewidth}{
\textbf{Finding 1}: Existing large language models and APR techniques fix very few Java vulnerabilities. Codex fixes 10.2 (20.4\%) vulnerabilities on average, exhibiting the best fixing
capability.
}}\smallskip

\subsubsection{LLMs Fine-Tuned with APR Data}
We applied LLMs fined-tuned with general APR data by \cite{clmapr23} on the vulnerability benchmarks. 
We cannot fine-tune Codex as OpenAI does not provide a public API for fine-tuning. \autoref{tab:repair_result} shows that all the fined-tuned LLMs fix more vulnerabilities than their original models. In detail, fine-tuned InCoder fixes 9 vulnerabilities, 4 more than its original model. The second best models is fine-tuned CodeGen, which fixes 8 vulnerabilities, 6 more than its original model. Fine-tuned CodeT5 and fine-tuned PLBART each fixes 3 and 2 more vulnerabilities.

Overall, fine-tuning with general APR data can improve the fixing capabilities of LLMs for vulnerabilities. First, fine-tuning could adapt LLMs to APR tasks better, making LLMs be aware of generating patches instead of open-ending code or text. Second, though vulnerabilities have special characteristics (root causes) compared to general bugs, some vulnerabilities still share similar repair patterns with general bugs, such as replacing a function argument with another variable, which can be well learned during fine-tuning. 
Given the scarcity of real-world vulnerability data, our results implicate that fine-tuning LLMs with general APR data 
can be beneficial.%

\smallskip\noindent\fbox{
\centering
\parbox{0.95\linewidth}{
\textbf{Finding 2}: Fine-tuning with general APR data  improves all four LLMs' vulnerability-fixing capabilities. Fine-tuned InCoder fixes 9 vulnerabilities, exhibiting competitive fixing capability compared to Codex's.}
}\smallskip

We also evaluate the compilation rates (i.e., portions of generated patches that compile) to study the quality of the patches. Uncompilable patches cannot be correct patches. Codex, the best model overall, has a compilation rate of 79.7\%, which is significantly higher than that of the best fine-tuned LLM, fine-tuned InCoder (55.2\%) and the best APR model, Recoder (57.6\%). 
Fine-tuning notably improves CodeT5 and CodeGen's compilation rates, from 6.4\% to 46.8\% and from 35.8\% to 47.2\% respectively.  On the other hand, the compilation rate of fine-tuned PLBART is 45.2\%, slightly lower than the original PLBART's compilation rate of 47.8\%. 
Despite the higher 65.2\% compilation rate of InCoder compared to its fine-tuned model,  it generates 82.0\% duplicate patches, whereas the fine-tuned InCoder generates patches with more diverse modifications that result in more correct fixes. %
Overall, compared with compilation rates of repairing general bugs~\cite{clmapr23}, these compilation rates of fixing vulnerability are lower. PLBART, CodeGen and InCoder without fine-tuning when repairing general bugs show an average of 65\%--73\% compilation rate~\cite{clmapr23}, outperforming both of their original and fine-tuned models when repairing vulnerabilities.

\begin{figure}[!t]
\centering

\begin{subfigure}[b]{0.47\textwidth}
   \vspace{2mm}
   \includegraphics[trim=0mm 0mm 0mm 0mm,clip,width=1\linewidth]{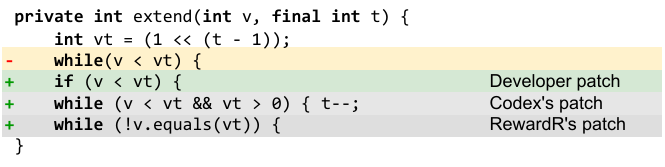}
   \caption{Vul4J-12 and its uncompilable patches}
   \label{fig:vul4j-12} 
\end{subfigure}

\begin{subfigure}[b]{0.47\textwidth}
   \centering
   \includegraphics[trim=0mm 0mm 0mm 0mm,clip,width=1\linewidth]{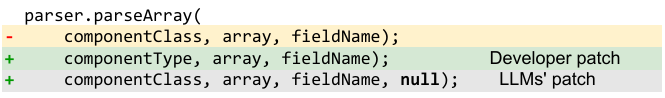}
   \caption{Vul4J-1 and its uncompilable patches}
   \label{fig:vul4j-1}
\end{subfigure}

\caption{Vul4J-12's and VulJ-1's developer patch and uncompilable patch}
\label{fig:uncompilable}
\end{figure}

\autoref{fig:vul4j-12} shows an example of uncompilable patches of Vul4J-12: The function signature declares \code{t} to be \code{final}, thus \code{t}'s value is not allowed to be changed. However, Codex fails to capture this constraint, even though the function signature is only two lines above the buggy line. As a result, it generates code \code{t\text{-}\text{-}} to decrease \code{t}'s value which makes the patch uncompilable. Similary, RewardR ignores the fact that \code{v} and \code{vt} are both of type \code{int}, and invokes the invalid function \code{equals} on them. \autoref{fig:vul4j-1} shows another example of uncomplable patch for Vul4J-1: \code{parseArray} is a method defined in another class in the project that accepts two or three arguments only. All the four fine-tuned LLMs generate the same uncompilable patches where they pass \code{null} as the fourth argument, because they do not have the information that \code{parseArray} does not accept four arguments. 

These results suggest that LLMs' abilities to learn code syntax could be improved. Recent work~\cite{jiang2021cure,zhu2021syntax} are steps in the right direction to add domain knowledge to models to help them learn code syntax and semantics. %
Another direction is prompt engineering, such as providing method signatures or type information in the prompt to specify the constraints. This would enable LLMs to utilize syntax information from across the entire project, rather than being limited to the code within the buggy function.

\smallskip\noindent\fbox{
\centering
\parbox{0.95\linewidth}{
\textbf{Finding 3}: Codex has the highest compilation rate of 79.7\%. 
Other LLMs (fine-tuned or not) and APR techniques have low compilation rates (the lowest of 6.4\% with CodeT5 and the rest between 24.5\% to 65.2\%), showing a lack of syntax domain knowledge.}
}\smallskip

\subsection{RQ2: What kinds of vulnerabilities do LLMs and learning-based APR techniques fix?}
\label{sec:rq2}

\begin{table*}[!t]

\centering
\footnotesize
\caption{
Detailed description of the vulnerabilities fixed by each LLM, fine-tuned LLM, and DL-based APR technique
}
\label{tab:uion_table} 

\begin{tabular}{r@{\quad}r@{\quad }l@{\enspace}r@{\enspace}r@{\enspace}r@{\enspace}r@{\enspace}r@{\quad}r@{\enspace}r@{\enspace}r@{\enspace}r@{\quad}r@{\enspace}r@{\enspace}r@{\enspace}r}
\toprule
 & & & \multicolumn{5}{c}{\textbf{LLMs}} & \multicolumn{4}{c}{\textbf{Fine-tuned LLMs}} & \multicolumn{4}{c}{\textbf{APR Techniques}} \\ %
\cmidrule(lr){4-8}\cmidrule(lr){9-12}\cmidrule(lr){13-16}
\textbf{Vul. ID} & \textbf{CWE} & \textbf{Description}
& \rotatebox[origin=l]{0}{Codex} 
& \rotatebox[origin=l]{0}{CodeT5}
& \rotatebox[origin=l]{0}{CodeGen}
& \rotatebox[origin=l]{0}{PLBART}
& \rotatebox[origin=l]{0}{InCoder}
& \rotatebox[origin=l]{0}{CodeT5}
& \rotatebox[origin=l]{0}{CodeGen}
& \rotatebox[origin=l]{0}{PLBART}
& \rotatebox[origin=l]{0}{InCoder}
& \rotatebox[origin=l]{0}{CURE}
& \rotatebox[origin=l]{0}{Recoder}
& \rotatebox[origin=l]{0}{RewardR}
& \rotatebox[origin=l]{0}{KNOD}
\\ %
\midrule
Vul4J-1   & 20  & Improper Input Validation    & \checkmark   &  &   &  &    &  &  \checkmark &  &  \checkmark &  &    &    &    \\
Vul4J-4   & \textit{unk}  &       /      & \checkmark   &  &  &  &   &  \checkmark &  \checkmark &  &  \checkmark &  &    &     &   \\
Vul4J-5   & \textit{unk} &    /    & \checkmark &  &   &   &   &   &   &  \checkmark&  \checkmark &  &   &   &     \\
Vul4J-12  & 835 & Infinite Loop       & \checkmark & \checkmark &   &   &   &   &  \checkmark & \checkmark &  \checkmark & \checkmark   &    &     &   \\
Vul4J-19  & \textit{unk} &   /  &  \checkmark   &  &   &  &    &  &   &  &   &  &    &    &    \\
Vul4J-20  & \textit{unk} &   /  &  \checkmark   &  &   &  &    &  &   &  &   &  &    &    &    \\
Vul4J-25  & 39 & Cross-site Scripting   & \checkmark &  &   &  &    &  &   &  &   &  &    &     &   \\
Vul4J-39  & 200 & Sensitive Information   &  & \checkmark & \checkmark  &  &  \checkmark &  \checkmark &  \checkmark &  &  \checkmark &  &    &     \\
Vul4J-47  & 611 & Improper External References &  \checkmark   &  &   &   &   &  &   &  &   &  &    &     &   \\
Vul4J-50  & 79  & Cross-site Scripting &  \checkmark   &  &   &   &  \checkmark &  &  \checkmark &  &  \checkmark &  &    &    &    \\
Vul4J-59  & 79  & Cross-site Scripting &  \checkmark   &  &   &   &  \checkmark &  &   &  &   &  &    &    &    \\
Vul4J-73  & 522 & Protected Credentials &  \checkmark  &  &   &  &    &  &   &  &   &  &    &    &  \checkmark \\
Halo-1    & 22  & Path Traversal   &  \checkmark   &  &   &   &  \checkmark &  &  \checkmark &  &  \checkmark &  &   &   &   \\
Jenkins-2 & 200 & Sensitive Information &  \checkmark   &   &   & \checkmark   &  \checkmark &  \checkmark &   &  &  \checkmark &  &    &  \checkmark  &     \\
Jenkins-3 & 200 & Sensitive Information  &  \checkmark   &  & \checkmark &  \checkmark &   &  \checkmark &  \checkmark & \checkmark &   &  &   &    &    \\ 
Ratpack-1 & 74 & Improper Neutralization &  \checkmark   &  &    &   &   &  \checkmark &  \checkmark & \checkmark &  \checkmark &  &  \checkmark  &  \checkmark  &    \\\hline{}
\#Total: 16 &  &   & 15 & 2 & 2 & 2 & 5 &  5 &  8 & 4  & 9  & 1&  1 &2  &  1  \\
\bottomrule
\end{tabular}
\end{table*}

Table~\ref{tab:uion_table} shows the vulnerabilities that are correctly fixed by the LLMs, fine-tuned LLMs, and APR techniques. 
In total, 16 vulnerabilities (belonging to ten CWE categories as shown in column \textit{CWE} with their description in column \textit{Description}) from both benchmarks  are fixed by at least one of the models. The IDs of these vulnerabilities are listed under column \textit{Vul. ID}. Some vulnerabilities belong to no specific CWE category and are listed as \textit{unk}.

\begin{figure}[!t]
\centering
\begin{subfigure}[b]{0.47\textwidth}
   \centering
   \includegraphics[width=1\linewidth]{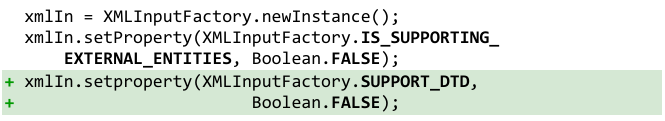}
   \caption{Vul4J-47 and its developer patch}
   \label{fig:vul4j-47-a}
\end{subfigure}   
\begin{subfigure}[b]{0.47\textwidth}
   \centering
   \includegraphics[width=1\linewidth]{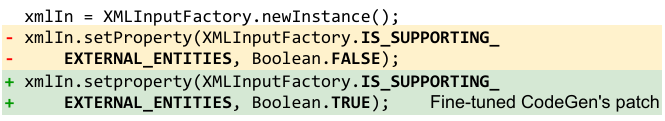}
   \caption{Vul4J-47 and the incorrect patch generated by fine-tuned CodeGen}
   \label{fig:vul4j-47-b}
\end{subfigure}   

\caption{Java vulnerability Vul4J-47 and its patches}
\label{fig:vul4j-47}
\end{figure}

Vul4J-47 is a vulnerability that only Codex can fix. \autoref{fig:vul4j-47-a} shows the developer patch for Vul4J-47 of type CWE-611 (Improper Restriction of XML External Entity Reference) and CWE-918 (Server-Side Request Forgery). The correct fix requires inserting a statement \code{xmlIn.setProperty(XMLInputFactory.
SUPPORT\_DTD, Boolean.FALSE)} to disable the support of Document Type Definition (DTD), because  DTD can be used to perform server-side request forgery (SSRF) attacks. %
The original buggy code only disables the support for external entities by setting the \code{IS\_SUPPORTING\_EXTERNAL\_
ENTITIES} property to false, which is not enough to prevent the attack.  \autoref{fig:vul4j-47-b} shows an incorrect patch generated by fine-tuned CodeGen, which replaces the \code{Boolean.FALSE} with \code{Boolean.TRUE}. In general, except Codex, other LLMs and fine-tuned LLMs  only fix vulnerabilities that require simple modifications such as deleting statements or replacing variable/method names.

On the other hand, Codex fixes 15 out of the 16 vulnerabilities (the union of all  bugs, for which Codex generates at least one correct patch in twenty-five runs).
The one vulnerability fixed by other LLMs but not Codex is Vul4J-39 of type CWE-200 (Exposure of Sensitive Information to an Unauthorized Actor). This vulnerability can be fixed by simply deleting the entire buggy code. However, for Vul4J-39, Codex generates patches by applying different modifications to the buggy code, rather than deleting it.

\begin{figure}[!t]
   \centering
   \includegraphics[width=1\linewidth]{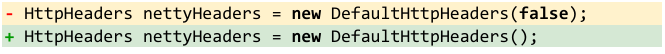}
   \caption{Java vulnerability Ratpack-1 and its developer patch}
   \label{fig:ratpack}
\end{figure}

Ratpack-1, Vul4J-12, Vul4J-39 and Jenkins-2 are four vulnerabilities fixed by the most number (6-7 out of 13) of models. Ratpack-1 (\autoref{fig:ratpack}) when initializing \code{DefaultHttpHeaders}, sets the constructor argument to \code{false}, which disables the validation for user-supplied header values. The correct patch is simply removing \code{false} or changing it to \code{true} to enable the validation. The fix for Vul4J-12 (\autoref{fig:vul4j-12}) is to change the keyword \code{while} to \code{if}, and the fix for both Jenkins-2 and Vul4J-39 is to simply delete of an \code{if} statement that exposes sensitive information to unauthorized actors. The simplicity of these patches are evident from the number of models that can fix them.

\smallskip\noindent\fbox{
\centering
\parbox{0.95\linewidth}{
\textbf{Finding 4}:  Large language models and APR techniques, except Codex,  only fix vulnerabilities that require simple changes, such as deleting statements or replacing variable/method names. %
}
}\smallskip

Surprisingly, the nine LLMs and {\numapr} APR techniques fix none of the six new CWE types that \bench adds, which shows that our \bench helps reveal the limitations of existing LLMs and APR techniques in fixing Java vulnerabilities. This calls for new techniques that can fix 
CWE-172, %
CWE-325, %
CWE-347,
CWE-444, %
CWE-668, %
and CWE-1295. %
In addition, for CWE-611 
that is covered by Vul4J's Vul4J-47, we add  two instances of this CWE type (Quartz-1 and Retrofit-1) in \bench. 
Codex fixes Vul4J-47, but %
none of the LLMs and APR techniques fixes the additional Quartz-1 and Retrofit-1. This shows that \bench complements Vul4J even on CWE categories that Vul4J has already covered.

\begin{figure}[!t]
\centering
\begin{subfigure}[b]{0.47\textwidth}
   \centering
   \includegraphics[trim=0mm 1mm 0mm 0mm,clip,width=1\linewidth]{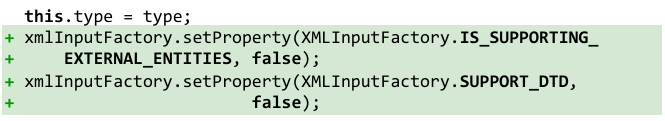}
   \caption{Developer patch of Retrofit-1}
   \label{fig:retrofit-a}
\end{subfigure}   
   
\begin{subfigure}[b]{0.47\textwidth}
   \vspace{2mm}
   \includegraphics[trim=0mm 1mm 0mm 0mm,clip,width=1\linewidth]{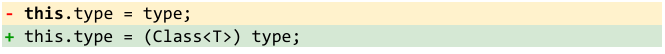}
   \caption{Incorrect fix generated by InCoder}
   \label{fig:retrofit-b} 
\end{subfigure}

\caption{Java vulnerability Retrofit-1 and its patches.}
\label{fig:retrofit}
\end{figure}

\begin{figure}[!t]
   \centering
   \includegraphics[width=1\linewidth]{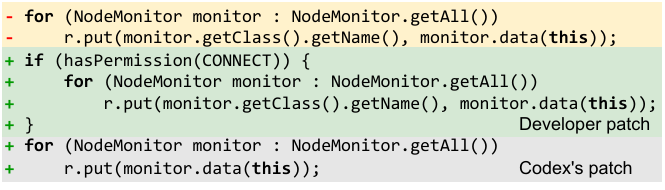}
   \caption{Java vulnerability Jenkins-1 and its patches}
   \label{fig:jenkins-1}
\end{figure}

\begin{table*}[!t]
\centering
\footnotesize
\caption{Impact of code transformation on LLMs' and APR models' vulnerability repair capabilities. For Codex, x $\pm$ y: x denotes the average number of correctly fixed bug, and y denotes the margin of error (95\% confidence).}
\label{tab:trans_repair_result}

\begin{tabular}
{r@{\quad}r@{\enspace}r@{\enspace}r@{\enspace}r@{\enspace}r@{\enspace}
r@{\quad}r@{\enspace}r@{\enspace}r@{\enspace}r@{\enspace}
r@{\quad}r@{\enspace}r@{\enspace}r@{\enspace}r@{\enspace}}
\hline
 & \multicolumn{5}{c}{LLMs}  &  & \multicolumn{4}{c}{Fine-Tuned LLMs} &  & \multicolumn{4}{c}{APR Techniques} \\ 
 \cline{2-6} \cline{8-11} \cline{13-16} 
    & \rotatebox[origin=l]{0}{Codex} 
& \rotatebox[origin=l]{0}{CodeT5}
& \rotatebox[origin=l]{0}{CodeGen}
& \rotatebox[origin=l]{0}{PLBART}
& \rotatebox[origin=l]{0}{InCoder}&

 & \rotatebox[origin=l]{0}{CodeT5}
& \rotatebox[origin=l]{0}{CodeGen}
& \rotatebox[origin=l]{0}{PLBART}
& \rotatebox[origin=l]{0}{InCoder}& 
 & \rotatebox[origin=l]{0}{CURE }
 & \rotatebox[origin=l]{0}{ Recoder }  
 & \rotatebox[origin=l]{0}{ RewardR}  
 & \rotatebox[origin=l]{0}{ KNOD}   \\ \hline
    No transformation      &  \textbf{10.2 $\pm$0.3} & 2  & 2   & 2  & \textbf{5}   &  & \textbf{5}  &  \textbf{8}   &  \textbf{4}  &  \textbf{9}  &  & 1  & 1     & 2  & 1     \\ 
Rename only    & \textbf{8.1 $\pm$0.3}  & 0  & 1   & 0  & \textbf{2}   &  &  \textbf{4}  & \textbf{6}   & \textbf{1}  & \textbf{7}   &  & 0  & 1     & 1  &  1  \\
Code structure change only & \textbf{10.0 $\pm$0.3}  & 0  & 2   & 2  & \textbf{1}   &  & \textbf{4}  & \textbf{6}   & \textbf{4}  & \textbf{7}   &  & 0  & 1     & 1    & 2   \\
Rename + code structure change &  \textbf{8.7 $\pm$0.4}  & 0  & 1   & 1  & \textbf{1}   &  & \textbf{3}  & \textbf{4}   & \textbf{3}  & \textbf{4}   &  & 0  & 1     & 1  & 0   \\

\hline
\end{tabular}

\end{table*}

\autoref{fig:retrofit} shows Retrofit-1 of CWE-611 category. 
None of the models fixes Retrofit-1. The correct patch is to prevent XML External Entity attacks by calling \code{xmlInputFactory.setProperty(...)} to disable the support for external entities and DTD. But as LLMs are not provided with 
information that the vulnerability is about XML External Entity attacks (as suggested by the CWE type), they  only make changes on the buggy code (\autoref{fig:retrofit-b}) unrelated to XML properties. 
\autoref{fig:jenkins-1} shows Jenkins-1 of CWE-325 (Missing cryptographic step), a new CWE category that \bench adds. The correct fix for the bug is adding if-condition to check the permission before the for-loop to restrict the access to \code{NodeMonitor}. As Codex's patch shown in \autoref{fig:jenkins-1}, all the models fail to fix the bug because they only apply general modifications to the \code{for}-loop and are unaware that the bug is related to the permission restriction. Further, the \code{hasPermission} method and the \code{CONNECT} variable are declared outside of the buggy function, thus the models have no knowledge about their usages. This reflects two problems for LLMs to fix Java vulnerabilities: (1) With only buggy lines pointed out, LLMs fail to generate patches targeting the vulnerability. This suggests that it is necessary to provide LLMs with more information about the vulnerability, such as CWE types. (2) More project-specific information is needed for LLMs to fix vulnerabilities, i.e., providing LLMs with related methods and variables declared outside of the buggy function.

\smallskip\noindent\fbox{
\centering
\parbox{0.95\linewidth}{
\textbf{Finding 5}: Our new \bench benchmark reveals that large language models and APR techniques fail to fix many CWE types, including CWE-172 (Encoding error), CWE-325 (Missing cryptographic step), 
CWE-444 (HTTP request smuggling), CWE-668 (Exposure of resource to wrong sphere), and CWE-1295 (Debug messages revealing unnecessary information).
}
}\smallskip

\subsection{
RQ3: Fixing Capabilities on Transformed Vulnerabilities}
\label{sec:rq3}

To mitigate the training-testing data overlapping threat, we apply code transformations to the benchmarks to study the generalization abilities of Codex and LLMs on unseen data (Section~\ref{obf}).
Table~\ref{tab:trans_repair_result} shows the number of vulnerabilities that LLMs as is, fine-tuned LLMs, and APR techniques can fix in four settings: (1) \textit{No transformation}---the original vulnerability dataset, (2) \textit{Rename only}---only identifier renaming is applied, (3) \textit{Code structure change only}---only code structure change is applied, and (4) \textit{Rename + code structure change}---both transformations are applied. 

Overall, code transformations make LLMs (fine-tuned or not) and APR techniques fix fewer vulnerabilities. For example, fine-tuned InCoder fixes nine vulnerabilities in Vul4J and \bench (no transformation), but only fixes four fully transformed vulnerabilities (Rename + Code structure change).
The impact of transformation is smaller on some models, e.g., Codex and fine-tuned CodeT5, demonstrating these models' robustness against code transformations and generalized learning capabilities. This result, to some extent, addresses the threat of Codex's non-public training data and reveals Codex's strong learning and vulnerability-fixing capability. Many models only fix two or fewer vulnerabilities without transformations, thus the impact of transformations cannot be big for these models. 
However, we see a general trend across almost all models that these code transformations make models fix fewer number of vulnerabilities.  

\begin{figure}[!t]
\centering
\begin{subfigure}[b]{0.47\textwidth}
   \centering
   \includegraphics[trim=0mm 1mm 0mm 0mm,clip,width=1\linewidth]{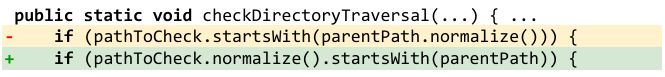}
   \caption{Halo's original buggy code and its correct patch}
   \label{fig:halo-1-a}
\end{subfigure}

\begin{subfigure}[b]{0.47\textwidth}
   \vspace{2mm}
   \includegraphics[trim=0mm 1mm 0mm 0mm,clip,width=1\linewidth]{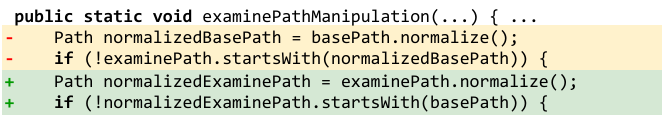}
   \caption{Halo's buggy code after Rename+Code structure change and its correct patch generated by Codex}
   \label{fig:halo-1-b} 
\end{subfigure}
\caption{Halo-1 before and after transformation } 
\end{figure}

\autoref{fig:halo-1-a} shows an example, Halo-1, whose correct fix is to call \code{normalize()} on \code{pathToCheck} to remove any redundant elements in the file path. This bug can be correctly fixed by Codex, fine-tuned CodeGen, and fine-tuned InCoder. Yet, after applying both transformations, only Codex can fix it (\autoref{fig:halo-1-b}). 

Different transformations have different effects but each transformation significantly affects at least one LLM. For example, although identifier renaming has small effect on fine-tuned CodeT5, it decreases the number of vulnerabilities that fine-tuned PLBART fixes by three. The result shows that our code transformation effectively tests the generalization ability of LLMs on unseen data.

\begin{figure}[!t]
\centering
\begin{subfigure}[b]{0.47\textwidth}
   \centering
   \includegraphics[trim=0mm 1mm 0mm 0mm,clip,width=1\linewidth]{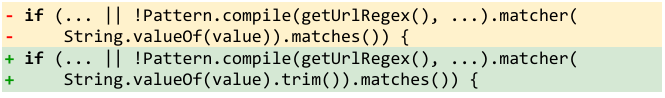}
   \caption{VUL4J-30's original buggy line and its correct patch}
   \label{fig:vul4j-30-a}
\end{subfigure}   
   
\begin{subfigure}[b]{0.47\textwidth}
   \vspace{2mm}
   \includegraphics[trim=0mm 1mm 0mm 0mm,clip,width=1\linewidth]{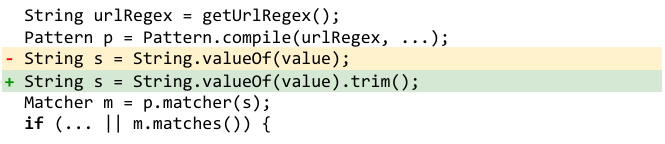}
   \caption{VUL4J-30's buggy line after code transformation and its correct patch generated by fine-tuned LLMs.}
   \label{fig:vul4j-30-b} 
\end{subfigure}

\caption{Vul4J-30 before and after code structure change}
\label{fig:vul4j-30}
\end{figure}

One interesting observation is that some models fix transformed vulnerabilities that they cannot fix in the original dataset. This is a reasonable phenomenon because our transformation may convert a code snippet into a simpler form for the models to fix. For example, Vul4J-30 is a bug that none of the models fixes in its  original form, but its transformed version is fixed by all  four fine-tuned LLMs when code structure transformation is applied. \autoref{fig:vul4j-30} shows that the fix of Vul4J-30 is to call \code{trim()} on \code{String.valueOf(value)}. The original vulnerability is hard to fix as \code{String.valueOf(value)} is a part of a complex if-condition. Yet, after code transformation, \code{String.valueOf(value)} stands out as a single statement, which is easier for LLMs to repair. This phenomenon suggests that equivalent code transformation could be a promising direction to simplify the vulnerable code and enhance the effectiveness of fixing vulnerabilities.

\smallskip\noindent\fbox{
\centering
\parbox{0.95\linewidth}{
\textbf{Finding 6}: Code transformations make large language models and APR techniques fix fewer number of vulnerabilities.  Some models such as Codex and fine-tuned CodeT5 are more robust to code transformations. On the other hand, some transformations make  vulnerabilities easier to fix. %
}
}\smallskip

\section{Threats to Validity}
\label{sec:threats}

Java vulnerabilities are diverse. It is hard for benchmarks to represent all of them. Thus, our findings might not generalize to all Java vulnerabilities. We address this threat by expanding the existing Java vulnerability benchmark with a new dataset of vulnerabilities. %

We rely on developers' patches to assess whether a vulnerability is fixed. Developers may make a mistake in  fixing vulnerabilities. Therefore, our ground truth might be incorrect. We mitigate this threat by only looking at vulnerabilities that are publicly disclosed in the NVD dataset that are reproducible and  include test cases indicating that the fixed version is no more exploitable.

Another threat is that Codex (and other LLMs) may have been trained on the vulnerability patches in Vul4J and \bench dataset. To mitigate this problem, we apply code transformations to create semantically equivalent vulnerabilities that are not included in their training dataset. Then we apply Codex to repair these transformed programs to prove that Codex is indeed able to repair new vulnerabilities that it has not seen. 

\section{Related Work}
\label{sec:related}
\subsection{DL-Based Vulnerability Fixing Techniques}
Much work uses DL to fix vulnerabilities. Encoder-decoder approaches have been proposed for repairing C vulnerabilities: \cite{fu2022vulrepair} fine-tuned a CodeT5 model with C vulnerability repair data; \cite{chen2022neural} trained a transformer model on a large bug fixing dataset and then tuned on a small vulnerability fixing dataset, but they use sequence accuracy as the evaluation metric rather than practical APR settings.
Previous work~\cite{huang2022repairing}
applied both CodeBERT and GraphCodeBert to fix vulnerabilities, but they 
only evaluated on a \emph{synthetic} vulnerability database, the Juliet 1.1 C/C++ test suite~\cite{boland2012juliet}, which is a benchmark for evaluating static analyzers only. As a result, the vulnerabilities in the dataset are isolated and simplified to fit within a few lines and are not representative of code vulnerabilities in the production.
Our work is different since we use a dataset of \emph{real-world} vulnerabilities for our evaluation, making our results closer to what researchers and developers can expect of the quality of LLM vulnerability repair in real-world production code. 

Prior work~\cite{pearce2022examining} applied LLMs with zero-shot learning to repair seven hand-crafted C/Python vulnerabilities and 12 real-world C vulnerabilities. They explored the effectiveness of different  prompt templates and used the static analysis tool CodeQL or C sanitizers to detect the vulnerabilities to incorporate the obtained error messages into the input prompts. Our work differs from~\cite{pearce2022examining} in several main aspects. First, we study not only LLMs but also DL-based APR tools and LLMs fine-tuned with general APR data. Second, we evaluate our approach on a larger dataset of 50 real-world Java vulnerabilities. Third, we apply code transformations to mitigate the data leakage problem and suggest a new direction of using transformations to simplify the repair for some vulnerabilities. %
Most vulnerabilities in Vul4J and VJBench cannot be detected by state-of-the-art Java security analysis tools, so we cannot incorporate error messages in the input prompts as~\cite{pearce2022examining} did.

\subsection{Vulnerability Benchmarks}
Previous work proposed benchmarks and datasets to help evaluate vulnerability fixing approaches. Maestro~\cite{pinconschi2022maestro} propose a platform for benchmarking tools on Java and C++ vulnerabilities. As Maestro does not support running LLMs and APR models,  we directly use the same Java vulnerability dataset, Vul4J~\cite{bui2022vul4j}, with our new dataset \bench.
Other benchmarks and datasets of real-world vulnerabilities have been proposed~\cite{ponta2019manually,bhandari2021cvefixes,fan2020ac,nikitopoulos2021crossvul}. However, these datasets only contain code snippets from the fixing commits and do not have test cases. Therefore, such datasets can only support code matching when evaluating the correctness of patches, and cannot be used in  automated program repair  in practice.

\subsection{LLMs for Repair and Other Tasks}
Researchers use LLMs to improve %
many software engineering tasks such as automated program repair~\cite{jiang2021cure,mashhadi2021applying,prenner2022can}, auto-complete suggestions~\cite{fan2022improving}, and pair-programming~\cite{imai2022github}. Much work also discusses the implication of LLMs for software developers~\cite{ernst2022ai,finnie2022robots,moroz2022potential} and current limitations of LLMs~\cite{sobieszek2022playing,dakhel2022github,asare2022github}. Our work explores a different application domain of LLMs, with its own challenges (vulnerabilities are notoriously difficult to fix~\cite{morrison2018vulnerabilities}) that have not been well explored yet.

\section{Conclusion}
\label{sec:conclusion}

This work is the first to investigate LLMs' and DL-based APR models' capacity at repairing vulnerabilities in Java.
We evaluate five LLMs, four fine-tuned LLMs,  and four DL-based APR techniques on two real-world Java vulnerability benchmarks including a new one that we create. 
We use code transformations to address the training and testing data overlapping threat of LLMs and create a new Java vulnerability repair benchmark \bench, and its transformed version \bench-trans. 
We find that existing LLMs and APR models fix very few Java vulnerabilities, and call for new research innovations to improve automated Java vulnerability repair such as creating larger vulnerability repair training datasets, fine-tuning LLMs with such data,  
exploring few-shot learning, and leveraging simplifying transformations to improve program repair.

 \smallskip\smallskip
\noindent
\textbf{Replication package}: Our benchmark and artifacts are available at~\cite{vjbenchgit}.

\section*{Acknowledgement}
We thank the reviewers for their insightful comments and
suggestions. This work was funded in part by NSF 1901242, NSF 2006688, J.P. Morgan AI Faculty Research Awards, and Meta/Facebook Research Awards. Any opinions, findings, and conclusions in this paper are those of the authors only and
do not necessarily reflect the views of our sponsors.

\balance
\bibliographystyle{ACM-Reference-Format}
\bibliography{main}
\end{document}